\documentclass[aps,prl,twocolumn,superscriptaddress,groupedaddress]{revtex4}  
\usepackage{graphicx}  
\usepackage{dcolumn}   
\usepackage{bm}        
\usepackage{amssymb}   

\begin{document}



\title{Study of baryon acoustic oscillations with SDSS DR13 data
and measurements of $\Omega_k$ and $\Omega_\textrm{DE}(a)$}
\author{B.~Hoeneisen} \affiliation{Universidad San Francisco de Quito, Quito, Ecuador}

\date{December 13, 2016}

\begin{abstract}
\noindent
We measure the baryon acoustic oscillation (BAO) observables
$\hat{d}_\alpha(z, z_c)$, $\hat{d}_z(z, z_c)$, and $\hat{d}_/(z, z_c)$
as a function of redshift $z$ in the range 0.1 to 0.7
with Sloan Digital Sky Survey (SDSS) data release DR13.
These observables are independent and satisfy a	consistency
relation that provides discrimination against miss-fits due to
background fluctuations.
From these measurements	and the	correlation angle $\theta_\textrm{MC}$
of fluctuations	of the Cosmic Microwave	Background (CMB) we obtain
$\Omega_k = -0.015 \pm 0.030$, 
$\Omega_{\textrm{DE}} + 2.2 \Omega_k = 0.717 \pm 0.004$ 
and $w_1 = 0.37 \pm 0.61$ for dark energy density allowed to vary as
$\Omega_{\textrm{DE}}(a) = \Omega_{\textrm{DE}} [ 1 + w_1 ( 1 - a)]$.
We present measurements of $\Omega_{\textrm{DE}}(a)$ at
six values of the expansion parameter $a$.
Fits with several scenarios and data sets are presented.
The data is consistent with space curvature parameter
$\Omega_k = 0$ and $\Omega_{\textrm{DE}}(a)$
constant. 
\end{abstract}

\maketitle

\section{Introduction}

Peaks in the density of the primordial universe 
are the sources of acoustic waves
of the tightly coupled plasma of photons, electrons, protons and helium
nuclei. These acoustic waves propagate a distance
$r'_S \approx 145$ Mpc until the time of recombination and 
decoupling $t_\textrm{dec}$ \cite{Eisenstein, PDG}. 
(All distances in this article are co-moving,
i.e. are referred to the present time $t_0$.)
The baryon acoustic oscillation (BAO) distance $r'_S$ corresponds
to the observed correlation angle $\theta_\textrm{MC}$ of 
fluctuations of the cosmic microwave background (CMB) \cite{PDG}.
Dark matter follows the BAO waves.
The results, well after decoupling,
for an initial point-like peak in the density,
are two concentric shells of overdensity of
radius $\approx 145$ Mpc and $\approx 18$ Mpc \cite{Eisenstein, BAO1, BAO2}. 
The inner spherical shell 
becomes reprocessed by the
hierarchical formation of galaxies \cite{BH}, while the outer shell is
unprocessed to better than 1\% \cite{BAO2, Seo}
(or even 0.1\% with corrections \cite{BAO2, Seo})
and therefore is an excellent standard ruler to measure
the expansion parameter $a(t)$ of the universe as a function
of time $t$. 
Histograms of galaxy-galaxy distances show an excess
in the approximate range $145 - 11$ Mpc to $145 + 11$ Mpc.
We denote by $d'_\textrm{BAO}$ the mean of this BAO signal.
We set $r'_S = d'_\textrm{BAO} f$, where $f$ is a correction
factor due to the peculiar motions of galaxies ($f$ depends
on the orientation of the galaxy pair with respect to the
line of sight).
Measurements of these BAO signals
are well established: see Refs. \cite{BAO1} and \cite{BAO2} for extensive
lists of early publications.

In this article we present studies of BAO
with Sloan Digital Sky Survey (SDSS) publicly released data DR13 \cite{DR13}.
The study has three parts:

(i) We measure the BAO observables
$\hat{d}_\alpha(z, z_c)$, $\hat{d}_z(z, z_c)$, and $\hat{d}_/(z, z_c)$ \cite{bh1}
in six bins of redshift $z$ from 0.1 to 0.7.
These observables are galaxy-galaxy correlation distances,
in units of $c/H_0$, of galaxy pairs respectively transverse to
the line of sight, along the line of sight, and in an interval
of angles with respect to
the line of sight, for a reference (fictitious) cosmology.

(ii) We measure the space curvature parameter $\Omega_k$ and
the dark energy density relative to the critical density 
$\Omega_\textrm{DE}(a)$
as a function of the expansion parameter $a$ with the
following BAO data used as an uncalibrated standard ruler:
$\hat{d}_\alpha(z, z_c)$, $\hat{d}_z(z, z_c)$, and $\hat{d}_/(z, z_c)$
for $0.1 < z < 0.7$ (this analysis), $\theta_\textrm{MC}$ for $z_\textrm{dec} = 1089.9 \pm 0.4$ from
Planck satellite observations \cite{PDG, Planck}, and
measurements of BAO distances in the Lyman-alpha (Ly$\alpha$) forest
with SDSS BOSS DR11 data at $z = 2.36$ \cite{lyman} and $z = 2.34$ \cite{lyman2}.

(iii) Finally we use the BAO measurements
as a calibrated standard ruler to
constrain a wider set of cosmological parameters.

The present analysis with DR13 data \cite{DR13} closely follows the methods
developed in Refs. \cite{bh1} and \cite{bh2} for data release DR12 \cite{DR12}.

\section{BAO observables}

To define the quantities being measured we
write the (generalized) Friedmann equation that describes the
expansion history of a homogeneous universe:
\begin{equation}
\frac{1}{H_0} \frac{1}{a} \frac{da}{dt} \equiv E(a) =
\sqrt{\frac{\Omega_\textrm{m}}{a^3} + \frac{\Omega_{\textrm{r}}}{a^4}
+ \Omega_\textrm{DE}(a) + \frac{\Omega_k}{a^2}}. 
\label{E}
\end{equation}
The expansion parameter $a(t)$ is normalized so that
$a(t_0) = 1$ at the present time $t_0$. 
The Hubble parameter $H_0 \equiv 100 h$ km s$^{-1}$ Mpc$^{-1}$ is normalized
so that $E(1) = 1$ at the present time, i.e.
\begin{equation}
\Omega_\textrm{m} + \Omega_\textrm{r} + \Omega_\textrm{DE}(1) + \Omega_k= 1.
\end{equation}
The terms under the square root in Eq. (\ref{E}) are 
densities relative to the critical density of,
respectively,
non-relativistic matter, ultra-relativistic radiation,
dark energy (whatever it is), and space curvature.
In the General Theory of Relativity
$\Omega_\textrm{DE}(a)$ is constant. Here we allow
$\Omega_\textrm{DE}(a)$ be a function of $a$ to be 
determined by observations.
Measuring $\Omega_k$ and $\Omega_\textrm{DE}(a)$
is equivalent to measuring the expansion history
of the universe $a(t)$. The expansion parameter $a$
is related to redshift $z$ by $a = 1/(1 + z)$.

The distance $d'$ between two galaxies observed with a
relative angle $\alpha$ and redshifts $z_1$ and $z_2$
can be written, with sufficient accuracy for our purposes, as \cite{bh1}
\begin{eqnarray}
d' & \equiv & \frac{c}{H_0} d, \nonumber \\
d & = & \sqrt{ d^2_\alpha + d^2_z }, \nonumber \\
d_\alpha & = & 2 \sin{\left( \frac{\alpha}{2} \right)} \sqrt{\chi(z_1) \chi(z_2)}
\left[ 1 + \frac{1}{6} \Omega_k \chi(z_1) \chi(z_2) \right], \nonumber \\
d_z & = & \chi(z_1) - \chi(z_2), \qquad \textrm{where} \nonumber \\
\chi(z) & \equiv & \int_0^z{\frac{dz'}{E(z')}}.
\label{dprime}
\end{eqnarray}
$d_\alpha$ and $d_z$ are the distance components, in units of $c/H_0$, 
transverse to
the line of sight and along the line of sight, respectively.
($\chi(z)$ should not be confused with the $\chi^2$ of fits.)
The difference between the approximation (\ref{dprime})
and the exact expression for $d'$, given by Eq. (3.19) of Ref. \cite{Peacock},
is negligible for two galaxies at the distance $d_\textrm{BAO}$:
the term of $d_\alpha$ proportional to $\Omega_k$ in Eq. (\ref{dprime})
changes by $0.1\%$ at $z = 0.7$.

We find the following approximations to
$\chi(z)$ and $1/E(z)$ valid in the range $0 \le z < 1$
with precision approximately $\pm 1\%$ 
for $z_c \approx 3.79$ \cite{bh1}:
\begin{equation}
\chi(z) \approx z \exp{\left( - \frac{z}{z_c} \right)}, 
\frac{1}{E(z)} \approx \left( 1 - \frac{z}{z_c} \right) \exp{\left( - \frac{z}{z_c} \right)}.
\label{chi_approx}
\end{equation}

Our strategy is as follows:
We consider galaxies with redshift in a given range 
$z_{\textrm{min}} < z < z_{\textrm{max}}$.
For each galaxy pair we calculate $d_\alpha(z, z_c)$, $d_z(z, z_c)$ and $d(z, z_c)$
with Eqs. (\ref{dprime}) \textit{with the approximation} (\ref{chi_approx}) 
and fill one of three histograms of $d(z, z_c)$ (with weights to
be discussed later)
depending on the ratio $d_z(z, z_c) / d_\alpha(z, z_c)$:
\begin{itemize}
\item
If $d_z(z, z_c) / d_\alpha(z, z_c) < 1/3$ fill a histogram of $d(z, z_c)$ that obtains
a BAO signal centered at $\hat{d}_\alpha(z, z_c)$. For this histogram,
$d^2_z(z, z_c)$ is a small correction relative to $d^2_\alpha(z, z_c)$ that is calculated
with sufficient accuracy with the approximation (\ref{chi_approx}),
i.e. an error less than 0.2\% on $\hat{d}_\alpha(z, z_c)$.
\item
If $d_\alpha(z, z_c) / d_z(z, z_c) < 1/3$ fill a second histogram of $d(z, z_c)$
that obtains a BAO signal centered at $\hat{d}_z(z, z_c)$. For this histogram,
$d^2_\alpha(z, z_c)$ is a small correction relative to $d^2_z(z, z_c)$ that is calculated
with sufficient accuracy with the approximation (\ref{chi_approx})
and $\Omega_k = 0$, i.e. an error less than 0.2\% on $\hat{d}_z(z, z_c)$.
\item
Else, fill a third histogram of $d(z, z_c)$ that obtains a 
BAO signal centered at $\hat{d}_/(z, z_c)$. 
\end{itemize}
Note that these three histograms have different galaxy pairs,
i.e. have independent signals and independent backgrounds.

The galaxy-galaxy correlation distance $d_\textrm{BAO}$, 
in units of $c/H_0$, is obtained from the BAO 
observables $\hat{d}_\alpha(z, z_c)$, $\hat{d}_z(z, z_c)$, or $\hat{d}_/(z, z_c)$
as follows:
\begin{eqnarray}
d_\textrm{BAO} & = & \hat{d}_\alpha (z, z_c) \frac{\chi(z) \left[ 1 + \frac{1}{6} \Omega_k \chi^2(z) \right]}
 {z \exp{(-z/z_c)}},  \label{da_rs} \\
d_\textrm{BAO} & = & \hat{d}_z (z, z_c) \frac{1}{(1 - z/z_c) \exp{(-z/z_c)} E(z)}, \label{dz_rs} \\
d_\textrm{BAO} & = & \hat{d}_/(z, z_c) \left(\frac{\chi(z) \left[ 1 + \frac{1}{6} \Omega_k \chi^2(z) \right]}
 {z \exp{(-z/z_c)}}\right)^{n/3} \nonumber \\
& & \times \left(\frac{1}{(1 - z/z_c) \exp{(-z/z_c)} E(z)}\right)^{1-n/3}. \label{d_rs}
\end{eqnarray}

A numerical analysis obtains $n = 1.70$ for $z = 0.2$, dropping to
$n = 1.66$ for $z = 0.8$ (in agreement with the method introduced in \cite{Eisenstein}
that obtains $n \approx 2$ when $d$ covers all angles).
The redshifts $z$ in Eqs. (\ref{da_rs}), (\ref{dz_rs}) and (\ref{d_rs})
correspond to the weighted mean of $z$ in the 
interval $z_{\textrm{min}}$ to $z_{\textrm{max}}$.
The fractions in Eqs. (\ref{da_rs}), (\ref{dz_rs}) and (\ref{d_rs})
are within $\approx 1\%$ of 1 for $z_c = 3.79$.
Note that the limits of $\hat{d}_\alpha (z, z_c)$ or $\hat{d}_z (z, z_c)$ or 
$\hat{d}_/(z, z_c)$ as $z \rightarrow 0$ are all equal to $d_\textrm{BAO}$.

The independent BAO observables
$\hat{d}_\alpha(z, z_c)$, $\hat{d}_z(z, z_c)$, and $\hat{d}_/(z, z_c)$
satisfy the consistency relation
\begin{equation}
Q \equiv \frac{\hat{d}_/(z, z_c)}{\hat{d}_\alpha(z, z_c)^{n/3}  \hat{d}_z(z, z_c)^{1 - n/3}} = 1.
\label{Q}
\end{equation}

The approximations in Eqs. (\ref{chi_approx})
obtain galaxy-galaxy correlation distances
$\hat{d}_\alpha(z, z_c)$, $\hat{d}_z(z, z_c)$, and
$\hat{d}_/(z, z_c)$ of a reference (fictitious) cosmology. 
We emphasize that these approximations are undone by Eqs. (\ref{da_rs}),
(\ref{dz_rs}), and (\ref{d_rs}) so in the end $d_\textrm{BAO}$
has the correct dependence on the cosmological
parameters which is different for Eqs. (\ref{da_rs}), 
(\ref{dz_rs}), and (\ref{d_rs}).

The BAO observables $\hat{d}_\alpha(z, z_c)$, 
$\hat{d}_z(z, z_c)$, and $\hat{d}_/(z, z_c)$
were chosen because (i) they are dimensionless,
(ii) they are independent,
(iii) they do not depend on any
cosmological parameter, 
(iv) they are almost independent of $z$ 
(for an optimized value of $r_c \approx 3.79$) so that
a large bin $z_\textrm{max} - z_\textrm{min}$ may be analyzed,
and (v) satisfy the consistency relation (\ref{Q}) 
which allows discrimination against fits that converge on
background fluctuations instead of the BAO signal.

It is observed that fluctuations in the CMB have a correlation angle \cite{PDG, Planck}
\begin{equation}
\theta_\textrm{MC} = 0.010410 \pm 0.000005,
\label{tMC}
\end{equation}
(we have chosen a measurement by the Planck collaboration with no input from BAO).
The extreme precision with which $\theta_\textrm{MC}$ is measured
makes it one of the primary parameters of cosmology.
The correlation distance $r_S$, in units of $c/H_0$, is obtained from
$\theta_\textrm{MC}$ as follows:
\begin{eqnarray}
r_S & = & 2 \sin{\left( \frac{1}{2} \theta_\textrm{MC} \right)} \chi(z_\textrm{dec}) \nonumber \\
& & \times \left[ 1 + \frac{1}{6} \Omega_k \chi^2(z_\textrm{dec}) 
+ \frac{1}{120} \Omega^2_k \chi^4(z_\textrm{dec})
\right] \label{thetaMC_rs}.
\label{rS2}
\end{eqnarray}
For $\chi(z_\textrm{dec})$ we do not neglect 
$\Omega_\textrm{r} \equiv \Omega_\gamma N_\textrm{eq}/2$ of photons
or neutrinos (we take $N_\textrm{eq} = 3.38$ \cite{PDG} 
corresponding to 3 neutrino flavors).

\section{Galaxy selection and data analysis}

We obtain the following data from the SDSS DR13 catalog \cite{DR13} for all objects
identified as galaxies that pass quality selection flags:
right ascension ra, declination dec, redshift $z$, redshift uncertainty $zErr$,
and the absolute value of the magnitude $r$.
We require a good measurement of redshift, i.e. $zErr < 0.001$.
The present study is limited 
to galaxies with right ascension in the range $110^0$ to $270^0$,
declination in the range $-5^0$ to $70^0$, and redshift in the
range $0.10$ to $0.70$.
The galactic plane 
divides this data set into two independent sub-sets:
the northern galactic cap (N) and the southern galactic cap (S)
defined by 
dec $\gtrless 27.0^0 - 17.0^0 \left[ (\textrm{ra} - 185.0^0)/(260.0^0 - 185.0^0) \right]^2$.

We calculate the absolute luminosity $F$ of galaxies relative
to the absolute luminosity of a galaxy with $r = 19.0$ at $z = 0.35$,
and calculate the corresponding magnitude $r_{35}$.
We consider galaxies with $17.0 < r_{35} < 23.0$ (G).
We define ``luminous galaxies" (LG) with, for example, $r_{35} < 19.2$, 
and ``clusters" (C).
Clusters C are based on a cluster finding algorithm that starts with LG's as seeds,
calculates the total absolute luminosity 
of all G's within a distance 0.006 (in units of $c/H_0$), and then selects
local maximums of these total absolute luminosities above a threshold,
e.g. $r_{35} < 16.6$.

A ``run" is defined by a range of redshifts $(z_\textrm{min}, z_\textrm{max})$,
a data set, and a definition of galaxy and ``center".
For each of 6 bins of redshift $z$ from 0.10 to 0.70, and
each of 5 offset bins of $z$ from 0.15 to 0.65, and for each
data set N or S, and for each choice of
galaxy-center G-G, G-LG, LG-LG, or G-C (with several
absolute luminosity cuts),
we fill histograms
of galaxy-center distances $d(z, z_c)$ and obtain the BAO distances
$\hat{d}_\alpha(z, z_c)$, $\hat{d}_z(z, z_c)$, and $\hat{d}_/(z, z_c)$
by fitting these histograms. 

\begin{figure}
\begin{center}
\scalebox{0.465} {\includegraphics{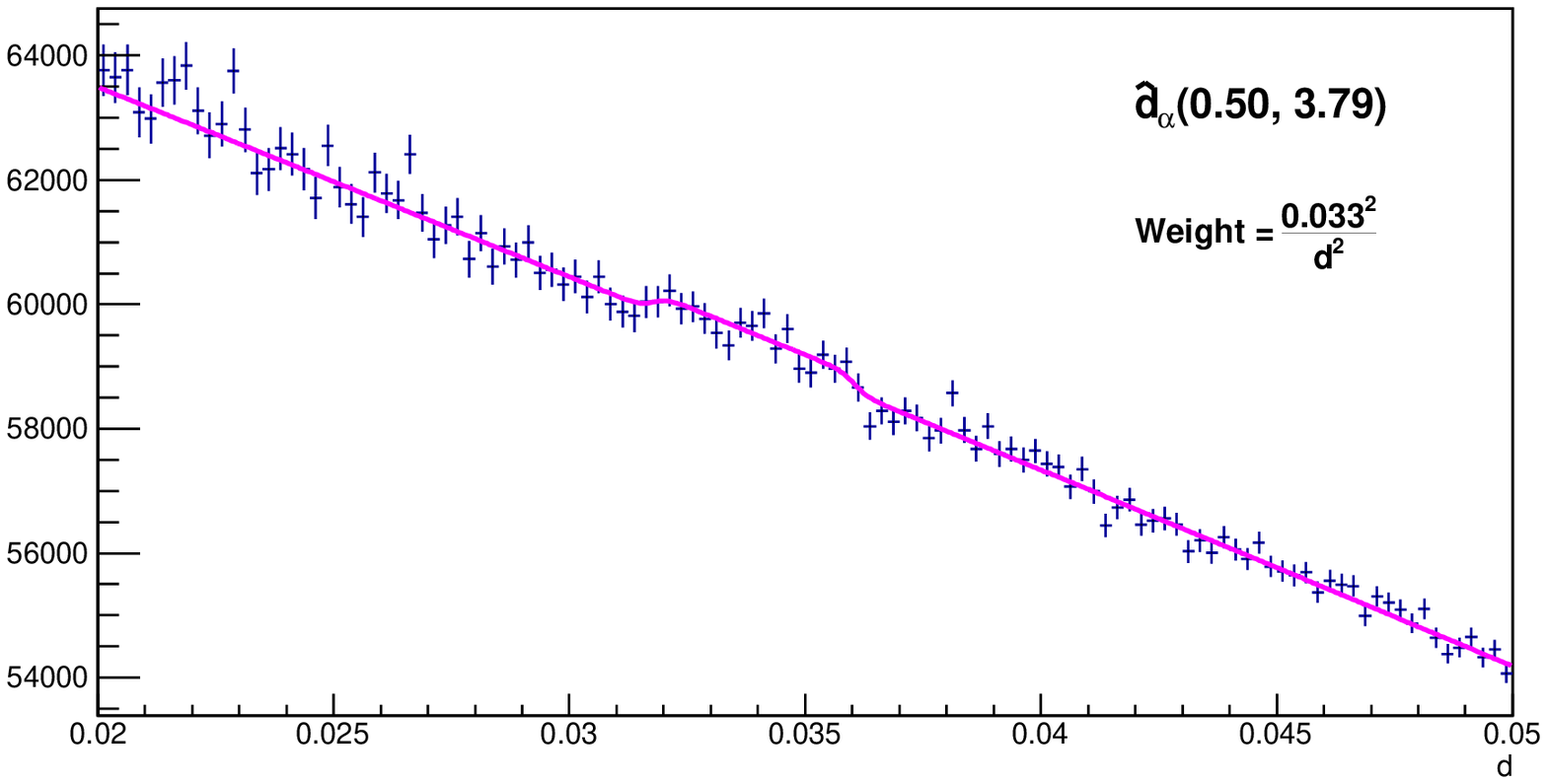}}
\scalebox{0.465} {\includegraphics{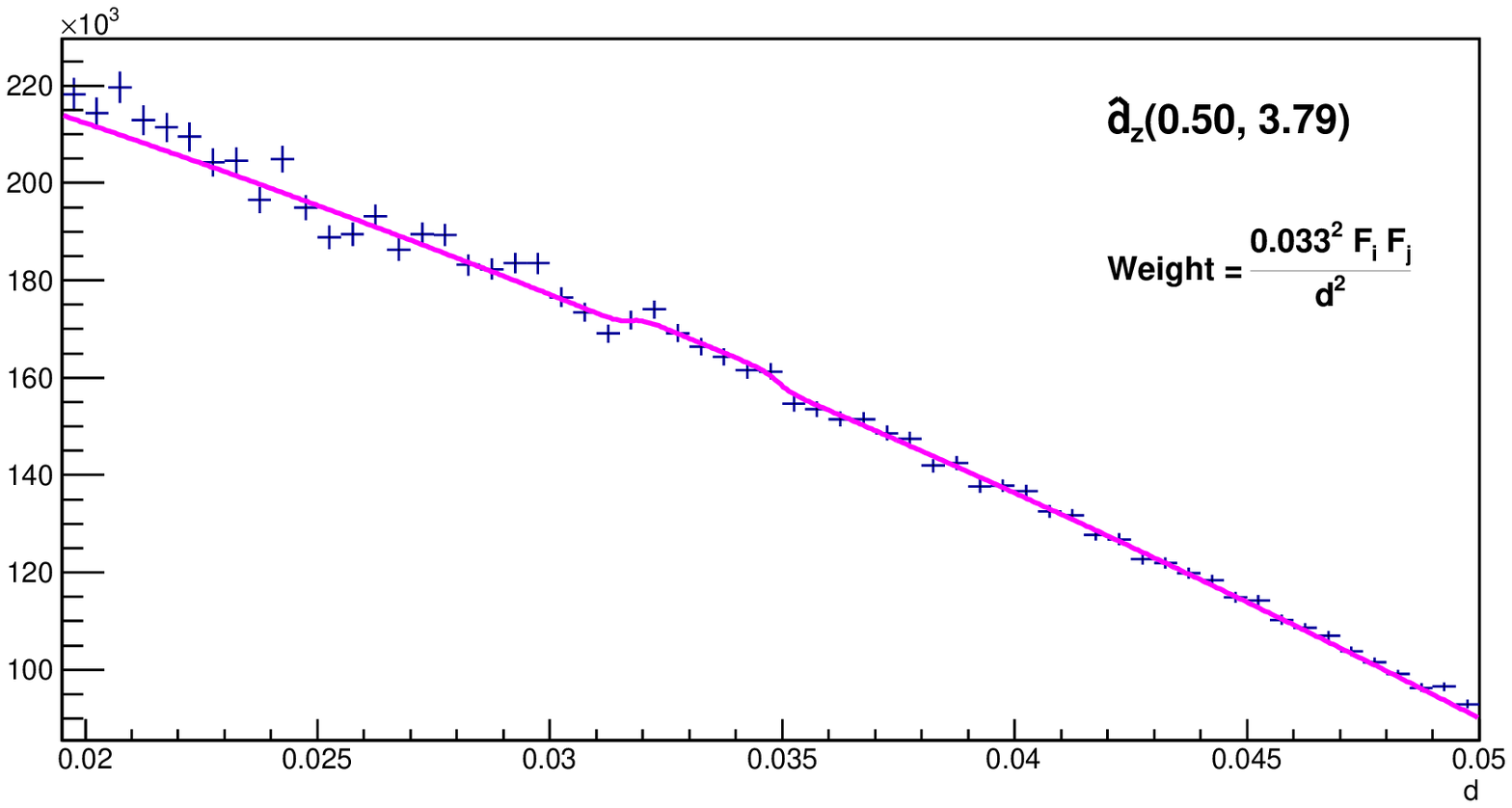}}
\scalebox{0.465} {\includegraphics{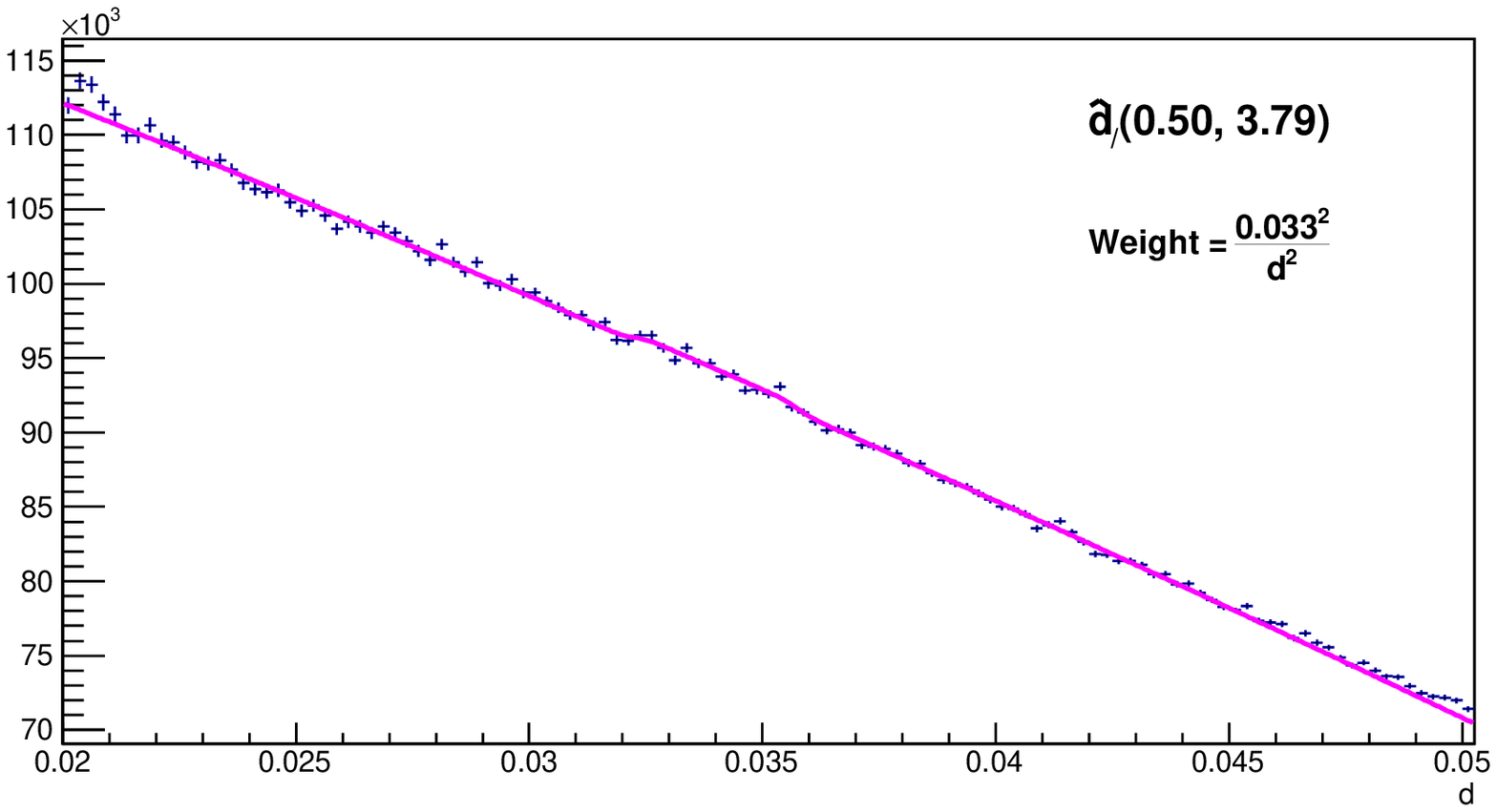}}
\caption{
Fits to histograms of $d(0.5, 3.79)$ that obtain
$\hat{d}_\alpha(0.5, 3.79)$, $\hat{d}_z(0.5, 3.79)$, and $\hat{d}_/(0.5, 3.79)$
in the northern galactic cap.
}
\label{fig_g_c_N_45_55}
\end{center}
\end{figure}

Histograms are filled with weights $(0.033/d)^2$
or $F_i F_j (0.033/d)^2$, where $F_i$ and $F_j$ are the absolute luminosities
$F$ of galaxy $i$ and center $j$ respectively.
We obtain histograms with $z_c = 3.79, 3.0$ and $5.0$.
The reason for this large degree of redundancy is the
difficulty to discriminate the BAO signal from the background with its
statistical and cosmological fluctuations due to galaxy clustering.
Pattern recognition is aided by multiple histograms with different
background fluctuations, and by the characteristic shape of
the BAO signal that has a lower edge at approximately
0.031 and an upper edge at approximately 0.036
as shown in Fig. \ref{fig_g_c_N_45_55}.

The fitting function is a second degree polynomial for the
background and, for the BAO signal, a step-up-step-down function
of the form
\begin{eqnarray*}
\frac{\exp{(x_<)}}{\exp{(x_<)} + \exp{(-x_<)}} - 
\frac{\exp{(x_>)}}{\exp{(x_>)} + \exp{(-x_>)}}
\end{eqnarray*}
where
\begin{eqnarray*}
x_< = \frac{d - \hat{d} + \Delta d}{\sigma}, \qquad 
x_> = \frac{d - \hat{d} - \Delta d}{\sigma}. 
\end{eqnarray*}

\begin{figure}
\begin{center}
\scalebox{0.465}
{\includegraphics{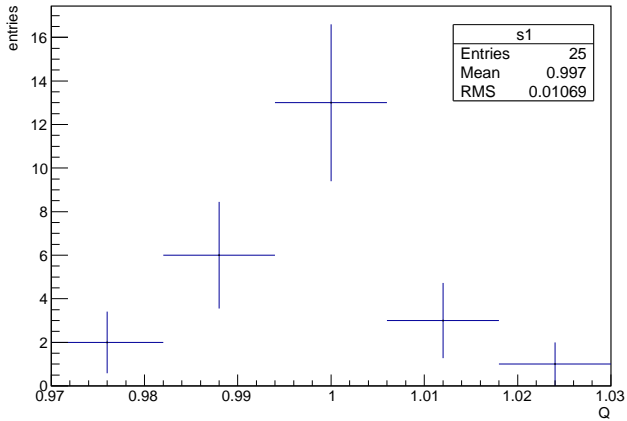}}
\caption{
Distribution of the consistency parameter $Q$ for
the 25 N or S successful runs.
}
\label{fig_Q}
\end{center}
\end{figure}

A run is defined as ``successful" if the fits \textit{to all
three histograms} converge with a signal-to-background ratio
significance greater than 1 standard deviation
(raising this cut further obtains little improvement
due to the cosmological fluctuations of the background),
and the consistency parameter $Q$ is in the
range 0.97 to 1.03 (if $Q$ is outside of this range then at
least one of the fits has converged on a fluctuation of the
background instead of the BAO signal).
We obtain 13 successful runs for N and 12 successful
runs for S which are presented in Tables \ref{N} and \ref{S}
respectively. The histogram of the consistency parameter $Q$
for these 25 runs is presented in Fig. \ref{fig_Q}.

For each bin of redshift $z$ we select from Tables \ref{N} or \ref{S} the run with
least $|Q - 1|$ and obtain the 18 independent BAO
distances listed in Table \ref{N_S}.
This Table \ref{N_S} is the main result of the present analysis, and
superceeds the corresponding tables for DR12 in Refs. \cite{bh1} and \cite{bh2}.

\section{Uncertainties}

Histograms of BAO distances $d(z, z_c)$ have 
statistical fluctuations, and 
fluctuations of the background
due to the clustering of galaxies 
as seen in Fig. \ref{fig_g_c_N_45_55}. These 
two types of fluctuations are
the dominant source of the total uncertainties of the
BAO distance measurements. These uncertainties are
independent for each entry in Table \ref{N_S}.
We present several estimates of the total uncertainties
of the entries in Tables \ref{N}, \ref{S}, and \ref{N_S} extracted 
directly from the fluctuations of the numbers in these tables.
All uncertainties in this article are at 68\% confidence level.

We neglect the variation of
$\hat{d}_\alpha(z, z_c)$, $\hat{d}_z(z, z_c)$, and $\hat{d}_/(z, z_c)$
between adjacent bins of $z$ with respect to their uncertainties.
The root-mean-square (r.m.s.) differences divided by $\sqrt{2}$ between corresponding
rows in Tables \ref{N} and \ref{S} for
$\hat{d}_\alpha(z, z_c)$, $\hat{d}_z(z, z_c)$, and $\hat{d}_/(z, z_c)$ are
0.00055, 0.00093, and 0.00054 respectively.
We assign these numbers as total uncertainties
of each entry in Tables \ref{N} and \ref{S}.

The 18 entries in Table \ref{N_S} are independent.
The r.m.s. differences for
rows 1-2, 3-4 and 5-6 divided by $\sqrt{2}$ are
0.00030, 0.00052, and 0.00020 for
$\hat{d}_\alpha(z, z_c)$, $\hat{d}_z(z, z_c)$, and $\hat{d}_/(z, z_c)$
respectively.

The average and standard deviation of the columns 
$\hat{d}_\alpha(z, z_c)$, $\hat{d}_z(z, z_c)$, and $\hat{d}_/(z, z_c)$
in Table \ref{N_S} are respectively
0.03342, 0.00021; 0.03355, 0.00051; and 0.03348, 0.00023.

The r.m.s. of $(1 - Q)$ for Tables N and S is 0.0111. 
The average of all entries in Tables N and S is 0.03383.
From the above estimates we take the uncertainties of
$\hat{d}_\alpha(z, z_c)$, $\hat{d}_z(z, z_c)$, and $\hat{d}_/(z, z_c)$
to be in the ratio $1:2:1$. 
From these numbers we calculate
the independent total uncertainties of
$\hat{d}_\alpha(z, z_c)$, $\hat{d}_z(z, z_c)$, and $\hat{d}_/(z, z_c)$
to be 0.00026, 0.00052, and 0.00026 respectively.

From these estimates, we take the following independent total
uncertainties for each entry of
$\hat{d}_\alpha(z, z_c)$, $\hat{d}_z(z, z_c)$, and $\hat{d}_/(z, z_c)$
in Table \ref{N_S}:
0.00030, 0.00060, and 0.00030 respectively.

\begin{table*}
\caption{\label{N}
Measured BAO distances $\hat{d}_\alpha(z, z_c)$,
$\hat{d}_z(z, z_c)$, and $\hat{d}_/(z, z_c)$ in units of $c/H_0$
with $z_c = 3.79$ (see text) from SDSS DR13 galaxies with
right ascension $110^0$ to $270^0$, and declination $-5^0$ to $70^0$
in the northern galactic cap, i.e. 
dec $> 27.0^0 - 17.0^0 \left[ (\textrm{ra} - 185.0^0)/(260.0^0 - 185.0^0) \right]^2$.
Uncertainties are statistical from the fits to the BAO signal.
Each BAO distance has an independent total uncertainty 
0.00055 for $\hat{d}_\alpha(z, z_c)$ and $\hat{d}_/(z, z_c)$, or
0.00093 for $\hat{d}_z(z, z_c)$.
No corrections have been applied.
}
\begin{ruledtabular}
\begin{tabular}{c|ccccc|ccc|c}
$z$ & $z_\textrm{min}$ & $z_\textrm{max}$ & galaxies & centers & type &
$100 \hat{d}_\alpha(z, z_c)$ & $100 \hat{d}_z(z, z_c)$ & $100 \hat{d}_/(z, z_c)$ & Q \\
\hline
$0.14$ & 0.10 & 0.20 & 152785 & 3729 & G-C & $3.367 \pm 0.014$ & $3.284 \pm 0.014$ & $3.306 \pm 0.018$ & 0.993 \\
$0.14$ & 0.10 & 0.20 & 152785 & 2853 & G-C & $3.376 \pm 0.017$ & $3.375 \pm 0.010$ & $3.363 \pm 0.017$ & 0.996 \\
$0.25$ & 0.20 & 0.30 & 58670 & 3271 & G-C & $3.271 \pm 0.019$ & $3.499 \pm 0.020$ & $3.414 \pm 0.020$ & 1.014 \\
$0.31$ & 0.25 & 0.35 & 69271 & 3677 & G-C & $3.344 \pm 0.025$ & $3.521 \pm 0.017$ & $3.409 \pm 0.016$ & 0.997 \\
$0.35$ & 0.30 & 0.40 & 83515 & 58491 & G-LG & $3.354 \pm 0.013$ & $3.320 \pm 0.010$ & $3.350 \pm 0.012$ & 1.003 \\
$0.40$ & 0.35 & 0.45 & 91672 & 4180 & G-C & $3.479 \pm 0.025$ & $3.358 \pm 0.028$ & $3.380 \pm 0.025$ & 0.987 \\
$0.46$ & 0.40 & 0.50 & 137972 & 137972 & G-G & $3.523 \pm 0.011$ & $3.420 \pm 0.019$ & $3.485 \pm 0.007$ & 1.002 \\
$0.46$ & 0.40 & 0.50 & 137972 & 55925 & G-LG & $3.450 \pm 0.014$ & $3.423 \pm 0.011$ & $3.456 \pm 0.007$ & 1.005 \\
$0.50$ & 0.45 & 0.55 & 195144 & 1514 & G-C & $3.436 \pm 0.024$ & $3.285 \pm 0.015$ & $3.431 \pm 0.012$ & 1.018 \\
$0.50$ & 0.45 & 0.55 & 195144 & 5935 & G-C & $3.391 \pm 0.019$ & $3.332 \pm 0.020$ & $3.401 \pm 0.014$ & 1.011 \\
$0.55$ & 0.50 & 0.60 & 188410 & 1105 & G-C & $3.338 \pm 0.021$ & $3.293 \pm 0.018$ & $3.300 \pm 0.026$ & 0.971 \\
$0.64$ & 0.60 & 0.70 & 81624 & 81624 & G-G & $3.368 \pm 0.014$ & $3.572 \pm 0.021$ & $3.439 \pm 0.012$ & 0.995 \\
$0.64$ & 0.60 & 0.70 & 81624 & 33982 & G-LG & $3.378 \pm 0.009$ & $3.586 \pm 0.016$ & $3.481 \pm 0.009$ & 1.004 \\
\end{tabular}
\end{ruledtabular}
\end{table*}

\begin{table*}
\caption{\label{S}
Measured BAO distances $\hat{d}_\alpha(z, z_c)$,
$\hat{d}_z(z, z_c)$, and $\hat{d}_/(z, z_c)$ in units of $c/H_0$
with $z_c = 3.79$ (see text) from SDSS DR13 galaxies with
right ascension $110^0$ to $270^0$, and declination $-5^0$ to $70^0$
in the southern galactic cap, i.e.
dec $< 27.0^0 - 17.0^0 \left[ (\textrm{ra} - 185.0^0)/(260.0^0 - 185.0^0) \right]^2$.
Uncertainties are statistical from the fits to the BAO signal.
Each BAO distance has an independent total uncertainty
0.00055 for $\hat{d}_\alpha(z, z_c)$ and $\hat{d}_/(z, z_c)$, or
0.00093 for $\hat{d}_z(z, z_c)$.
No corrections have been applied.
}
\begin{ruledtabular}
\begin{tabular}{c|ccccc|ccc|c}
$z$ & $z_\textrm{min}$ & $z_\textrm{max}$ & galaxies & centers & type &
$100 \hat{d}_\alpha(z, z_c)$ & $100 \hat{d}_z(z, z_c)$ & $100 \hat{d}_/(z, z_c)$ & Q \\
\hline
$0.19$ & 0.15 & 0.25 & 58381 & 1538 & G-C & $3.338 \pm 0.024$ & $3.315 \pm 0.037$ & $3.321 \pm 0.011$ & 0.998 \\
$0.25$ & 0.20 & 0.30 & 38931 & 3865 & G-C & $3.311 \pm 0.035$ & $3.381 \pm 0.016$ & $3.360 \pm 0.013$ & 1.006 \\
$0.31$ & 0.25 & 0.35 & 46916 & 2828 & G-C & $3.277 \pm 0.017$ & $3.438 \pm 0.027$ & $3.333 \pm 0.018$ & 0.996 \\
$0.31$ & 0.25 & 0.35 & 46916 & 2559 & G-C & $3.271 \pm 0.028$ & $3.309 \pm 0.042$ & $3.295 \pm 0.036$ & 1.002 \\
$0.46$ & 0.40 & 0.50 & 91599 & 91599 & G-G & $3.319 \pm 0.013$ & $3.509 \pm 0.022$ & $3.366 \pm 0.016$ & 0.990 \\
$0.46$ & 0.40 & 0.50 & 91599 & 37456 & G-LG & $3.323 \pm 0.019$ & $3.445 \pm 0.024$ & $3.369 \pm 0.016$ & 0.998 \\
$0.46$ & 0.40 & 0.50 & 37456 & 37456 & LG-LG & $3.390 \pm 0.019$ & $3.536 \pm 0.013$ & $3.367 \pm 0.014$ & 0.975 \\
$0.64$ & 0.60 & 0.70 & 53518 & 53518 & G-G & $3.424 \pm 0.015$ & $3.327 \pm 0.019$ & $3.373 \pm 0.016$ & 0.997 \\
$0.64$ & 0.60 & 0.70 & 53518 & 23384 & G-LG & $3.393 \pm 0.017$ & $3.427 \pm 0.028$ & $3.373 \pm 0.019$ & 0.990 \\
$0.64$ & 0.60 & 0.70 & 53518 & 941 & G-C & $3.349 \pm 0.014$ & $3.316 \pm 0.046$ & $3.346 \pm 0.020$ & 1.003 \\
$0.64$ & 0.60 & 0.70 & 23384 & 23384 & LG-LG & $3.446 \pm 0.016$ & $3.371 \pm 0.026$ & $3.381 \pm 0.015$ & 0.991 \\
$0.64$ & 0.60 & 0.70 & 53518 & 689 & G-C & $3.416 \pm 0.020$ & $3.317 \pm 0.021$ & $3.316 \pm 0.036$ & 0.983 \\
\end{tabular}
\end{ruledtabular}
\end{table*}

\begin{table}
\caption{\label{N_S}
Independent measured BAO distances $\hat{d}_\alpha(z, z_c)$,
$\hat{d}_z(z, z_c)$, and $\hat{d}_/(z, z_c)$ in units of $c/H_0$
with $z_c = 3.79$ (see text) obtained by selecting, for each bin of $z$, the entry
with least $|Q - 1|$ in Tables \ref{N} or \ref{S}.
Each BAO distance has an independent total uncertainty
0.00030 for $\hat{d}_\alpha(z, z_c)$ and $\hat{d}_/(z, z_c)$, or
0.00060 for $\hat{d}_z(z, z_c)$.
No corrections have been applied.
}
\begin{ruledtabular}
\begin{tabular}{c|cc|ccc}
$z$ & $z_\textrm{min}$ & $z_\textrm{max}$ &
$100 \hat{d}_\alpha(z, z_c)$ & $100 \hat{d}_z(z, z_c)$ & $100 \hat{d}_/(z, z_c)$ \\
\hline
$0.14$ & 0.1 & 0.2 & $3.376$ & $3.375$ & $3.363$ \\
$0.25$ & 0.2 & 0.3 & $3.311$ & $3.381$ & $3.360$ \\
$0.35$ & 0.3 & 0.4 & $3.354$ & $3.320$ & $3.350$ \\
$0.46$ & 0.4 & 0.5 & $3.323$ & $3.445$ & $3.369$ \\
$0.55$ & 0.5 & 0.6 & $3.338$ & $3.293$ & $3.300$ \\
$0.64$ & 0.6 & 0.7 & $3.349$ & $3.316$ & $3.346$ \\
\end{tabular}
\end{ruledtabular}
\end{table}

\section{Corrections}

Let us consider	corrections to the BAO distances 
due to peculiar	velocities and peculiar displacements
of galaxies towards their centers.
A relative peculiar velocity $v_p$ towards the center
causes a reduction of the BAO distances
$\hat{d}_\alpha(z, z_c)$, $\hat{d}_z(z, z_c)$, and $\hat{d}_/(z, z_c)$
of order $0.5 v_p/c$. In addition, the Doppler shift
produces an apparent shortening of $\hat{d}_z(z, z_c)$
by $v_p/c$, and somewhat less for $\hat{d}_/(z, z_c)$.

We multiply the measured BAO distances
$\hat{d}_\alpha(z, z_c)$, $\hat{d}_z(z, z_c)$, and $\hat{d}_/(z, z_c)$
by correction factors $f_\alpha$, $f_z$ and $f_/$ respectively.
Simulations in Ref. \cite{Seo} obtain
$f_\alpha - 1 = 0.2283 \pm 0.0609 \%$ and $f_z - 1 = 0.2661 \pm 0.0820 \%$
at $z = 0.3$, 
$f_\alpha - 1 = 0.1286 \pm 0.0425 \%$ and $f_z - 1 = 0.1585 \pm 0.0611 \%$
at $z = 1$, and
$f_\alpha - 1 = 0.0435 \pm 0.0293 \%$ and $f_z - 1 = 0.0582 \pm 0.0402 \%$
at $z = 3$.
In the following sections we present fits with the
corrections
\begin{eqnarray}
f_\alpha - 1 & = & 0.00320 \cdot a^{1.35}, \nonumber \\ 
f_z - 1 & = & 0.00381 \cdot a^{1.35}, \nonumber \\
f_/ - 1 & = & 0.00350 \cdot a^{1.35}.
\label{correction}
\end{eqnarray}

The effect of these corrections can be seen by comparing
the first two fits in Table \ref{BAO_fit} below. 
An order-of-magnitude estimate of this correction
can be obtained by calculating the r.m.s. 
$v_p$ corresponding to modes with 
$k \equiv 2 \pi / \lambda < 2 \pi / (4 d'_\textrm{BAO})$
with Eq. (11) of Ref. \cite{BH} and normalizing the result to 
$\sigma_8$, i.e. to the r.m.s. density fluctuation in a volume
$(8 \textrm{Mpc}/h)^3$.

\begin{table*}
\caption{\label{BAO_fit}
Cosmological parameters obtained from the 18 independent BAO measurements in Table \ref{N_S}
in several scenarios. Corrections for peculiar motions are given by
Eq. (\ref{correction}) except, for comparison, the fit ``1*" which has no correction.
Scenario 1 has $\Omega_\textrm{DE}(a)$ constant.
Scenario 3 has $w = w_0$.
Scenario 4 has $\Omega_\textrm{DE}(a) = \Omega_\textrm{DE} \left[1 + w_1 (1 - a)\right]$.
}
\begin{ruledtabular}
\begin{tabular}{c|cccccc} 
   & Scenario 1* & Scenario 1 & Scenario 1 & Scenario 3 & Scenario 4 & Scenario 4 \\
\hline
$\Omega_k$ & $0$ fixed  & $0$ fixed  & $0.173 \pm 0.173$ & $0$ fixed & $0$ fixed 
  & $0.151 \pm 0.185$ \\
$\Omega_\textrm{DE} + 0.5 \Omega_k$ & $0.714 \pm 0.014$ & $0.716 \pm 0.014$ & $0.710 \pm 0.016$ 
  & $0.772 \pm 0.094$ & $0.749 \pm 0.049$ & $0.732 \pm 0.052$ \\
$w_0$ & n.a. & n.a. & n.a. & $-0.84 \pm 0.22$ & n.a. & n.a. \\
$w_1$ & n.a. & n.a. & n.a. & n.a. & $0.44 \pm 0.60$ & $0.33 \pm 0.70$ \\
$100 d_\textrm{BAO}$ & $3.38 \pm 0.02$ & $3.39 \pm 0.02$ & $3.38 \pm 0.03$ 
  & $3.36 \pm 0.05$ & $3.36 \pm 0.05$ & $3.36 \pm 0.05$ \\
$\chi^2/$d.f. & $10.9/16$ & $11.2/16$ & $10.2/15$ & $10.6/15$ & $10.7/15$ & $10.0/14$ \\
\end{tabular}
\end{ruledtabular}
\end{table*}

\begin{table*}
\caption{\label{BAO_thetaMC_fit_18}
Cosmological parameters obtained from the 18 BAO measurements in Table \ref{N_S}
plus $\theta_\textrm{MC}$ from Eq. (\ref{tMC}) in several scenarios.
Corrections for peculiar motions are given by
Eq. (\ref{correction}).
Scenario 1 has $\Omega_\textrm{DE}(a)$ constant.
Scenario 2 has $w(a) = w_0 + w_a (1 - a)$.
Scenario 3 has $w = w_0$.
Scenario 4 has $\Omega_\textrm{DE}(a) = \Omega_\textrm{DE} \left[1 + w_1 (1 - a)\right]$.
}
\begin{ruledtabular}
\begin{tabular}{c|cccccc} 
   & Scenario 1 & Scenario 1 & Scenario 2 & Scenario 3 & Scenario 4 & Scenario 4 \\
\hline
$\Omega_k$ & $0$ fixed  & $0.002 \pm 0.007$ & $0$ fixed  & 
  $0$ fixed & $0$ fixed & $-0.015 \pm 0.030$ \\
$\Omega_\textrm{DE} + 2.2 \Omega_k$ & $0.719 \pm 0.003$ & $0.718 \pm 0.004$ & $0.708 \pm 0.015$ 
  & $0.718 \pm 0.004$ & $0.718 \pm 0.004$ & $0.717 \pm 0.004$ \\
$w_0$ & n.a. & n.a. & $-0.87 \pm 0.19$ & $-0.99 \pm 0.04$ & n.a. & n.a. \\
$w_a$ or $w_1$ & n.a. & n.a. & $-0.60 \pm 0.93$ & n.a. & $0.06 \pm 0.15$ & $0.37 \pm 0.61$ \\
$100 d_\textrm{BAO}$ & $3.40 \pm 0.02$ & $3.39 \pm 0.02$ & $3.36 \pm 0.06$ 
  & $3.39 \pm 0.03$ & $3.39 \pm 0.03$ & $3.37 \pm 0.05$ \\
$\chi^2/$d.f. & $11.2/17$ & $11.2/16$ & $10.7/15$ & $11.1/16$ & $11.1/16$ & $10.8/15$ \\
\end{tabular}
\end{ruledtabular}
\end{table*}

\section{Measurements of $\Omega_k$ and $\Omega_\textrm{DE}(a)$
from uncalibrated BAO}

We consider five scenarios:
\begin{enumerate}
\item
The observed acceleration of the expansion of the universe is due
to the cosmological constant, i.e. $\Omega_\textrm{DA}(a)$ 
is constant.
\item
The observed acceleration of the expansion of the universe is due
to a gas of negative pressure with an equation of state $w \equiv p/\rho < 0$.
We allow the index $w$ be a function of $a$ \cite{BAO1, Chevallier, Linder}:
$w(a) = w_0 + w_a (1 - a)$. While this gas dominates $E(a)$ the
equation \cite{PDG}
\begin{equation}
\frac{d \rho}{dt} = -3 \frac{da/dt}{a} ( \rho + p )
\end{equation}
can be integrated with the result \cite{BAO1, Chevallier, Linder}
\begin{equation}
\Omega_\textrm{DE}(a) = \Omega_\textrm{DE} a^{-3(1 + w_0 + w_a)} \exp{\left\{ -3w_a (1 - a) \right\}}.
\end{equation}
If $w_0 = -1$ and $w_a = 0$ we obtain constant $\Omega_\textrm{DA}(a)$ 
as in the General Theory of Relativity.
\item
Same as Scenario 2 with $w(a)$ constant, i.e. $w_a = 0$.
\item
We assume $\Omega_\textrm{DA}(a) = \Omega_\textrm{DA} [1 + w_1 (1 - a)]$.
\item
$\Omega_\textrm{DA}(a)$ is arbitrary and needs to be measured 
at every $a$.
\end{enumerate}
Note that BAO measurements can constrain $\Omega_\textrm{DE}(a)$
for $0.3 \lesssim a \le 1$ where
$\Omega_\textrm{DE}(a)$ contributes significantly to $E(a)$.

Let us try to understand qualitatively how 
the BAO distance measurements presented in Table \ref{N_S}
constrain the cosmological parameters.
In the limit $z \rightarrow 0$ we obtain
$d_\textrm{BAO} = \hat{d}_\alpha(0, z_c) = \hat{d}_z(0, z_c) = \hat{d}_/(0, z_c)$,
so the first row with $z = 0.14$ in Table \ref{N_S}
approximately determines $d_\textrm{BAO}$.
This $d_\textrm{BAO}$ and
the measurement of, for example, $\hat{d}_z(0.3, z_c)$
then constrains the derivative of 
$\Omega_\textrm{m}/a^3 + \Omega_\textrm{DE} + \Omega_\textrm{k}/a^2$
with respect to $a$ at $z \approx 0.3$, i.e.
constrains approximately 
$\Omega_\textrm{DE} + 0.5 \Omega_k$.
We need an additional constraint for Scenario 1.
$d_\textrm{BAO}$ and $\theta_\textrm{MC}$ constrain the last two 
factors in Eq. (\ref{rS2}), i.e. approximately constrain
$\Omega_\textrm{DE} + 2.1 \Omega_k$.
The additional BAO distance measurements in Table \ref{N_S}
then also constrain $w_0$ and $w_a$, or $w_1$.

In Table \ref{BAO_fit} we present the
cosmological parameters obtained by minimizing the
$\chi^2$ with 18 terms corresponding to the
18 independent BAO distance measurements in Table \ref{N_S} for
several scenarios. We find that the data is in
agreement with the simplest cosmology with
$\Omega_k = 0$ and $\Omega_\textrm{DE}(a)$ constant
with $\chi^2$ per degree of freedom (d.f.) $11.2/16$, so no additional
parameter is needed to obtain a good fit to this data.
For free $\Omega_k$ we obtain
$\Omega_\textrm{DE} + 0.5 \Omega_k = 0.710 \pm 0.016$
for constant $\Omega_\textrm{DE}(a)$, or $0.732 \pm 0.052$
if $\Omega_\textrm{DE}(a)$ is allowed to
depend on $a$ as in Scenario 4.
We present the variable $\Omega_\textrm{DE} + 0.5 \Omega_k$
instead of $\Omega_\textrm{DE}$ because it has a smaller 
uncertainty.
The constraints on $\Omega_k$ are weak.

In Table \ref{BAO_thetaMC_fit_18} we present the
cosmological parameters obtained by minimizing the
$\chi^2$ with 19 terms corresponding to the
18 BAO distance measurements listed
in Table \ref{N_S}
plus the measurement of the correlation angle $\theta_\textrm{MC}$
of the CMB given in Eq. (\ref{tMC}).
We present the variable $\Omega_\textrm{DE} + 2.2 \Omega_\textrm{k}$
instead of $\Omega_\textrm{DE}$ because it has a smaller uncertainty.
We obtain 
\begin{eqnarray}
\Omega_k & = & -0.015 \pm 0.030, \nonumber \\
\Omega_\textrm{DE} + 2.2 \Omega_\textrm{k} & = & 0.717 \pm 0.004, \nonumber \\
w_1 & = & 0.37 \pm 0.61,
\label{19_terms}
\end{eqnarray}
when $\Omega_\textrm{DE}(a)$ is allowed to vary as in Scenario 4.
There is no tension between the data and 
the case $\Omega_k = 0$ and constant $\Omega_\textrm{DE}(a)$:
with these two constraints
we obtain $\Omega_\textrm{DE} = 0.719 \pm 0.003$
with $\chi^2/\textrm{d.f.} = 11.2/17$. 

We now add BAO measurements with SDSS BOSS DR11 data of quasar Ly$\alpha$ forest 
cross-correlation 
at $z = 2.36$ \cite{lyman} and Ly$\alpha$ forest autocorrelation at $z = 2.34$ \cite{lyman2}.
From the combination in Fig. 13 of Ref. \cite{lyman2} we obtain
\begin{eqnarray}
\frac{\chi(2.34) (1 + \Omega_k \chi^2(2.34)/6)}{d_\textrm{BAO} (1 + 2.34)} 
 & = & 10.95 \pm 0.36, \nonumber \\
\frac{1}{E(z = 2.34) d_\textrm{BAO}} & = & 9.14 \pm 0.20.
\label{Lyalpha}
\end{eqnarray}
From the 18 BAO plus $\theta_\textrm{MC}$ plus 2 Ly$\alpha$ measurements,
for free $\Omega_k$, and $\Omega_\textrm{DE}(a)$ allowed to vary as in Scenario 4,
we obtain $\Omega_k = -0.013 \pm 0.009$,
$\Omega_\textrm{DE} + 2.2 \Omega_\textrm{k} = 0.717 \pm 0.004$,
and $w_1 = 0.34 \pm 0.24$.
The $\chi^2$/d.f. is 17.6/17. 
Note that the Ly$\alpha$ measurements reduce the uncertainties
of $\Omega_k$ and $w_1$.
Requiring $\Omega_k = 0$ and
$\Omega_\textrm{DE}(a)$ constant raises the $\chi^2$/d.f. to 19.7/19,
so we observe no tension between the data and these two requirements,
and obtain $\Omega_\textrm{DE} + 2.2 \Omega_\textrm{k} = 0.719 \pm 0.003$.

\section{Detailed measurement of $\Omega_\textrm{DE}(a)$}

\begin{figure}
\begin{center}
\scalebox{0.45}
{\includegraphics{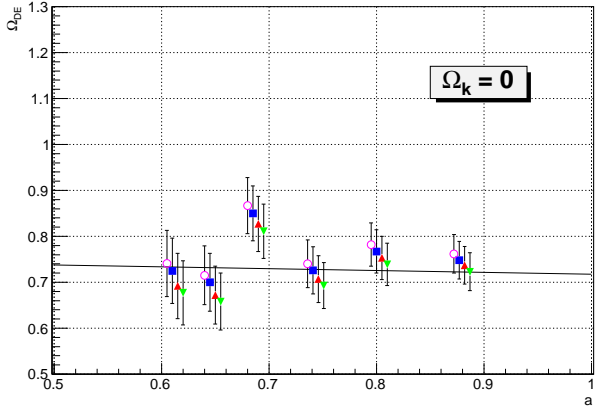}}
\caption{Measurements of
$\Omega_\textrm{DE}(a)$ obtained
from the 6 $\hat{d}_z(z, z_c)$ in Table \ref{N_S} 
for $\Omega_k = 0$, and
the corresponding $d_\textrm{BAO}$ and $\Omega_\textrm{DE}$
from the fit for Scenario 4 in Table \ref{BAO_thetaMC_fit_18}.
The straight line is 
$\Omega_\textrm{DE}(a) = 0.718 \left[ 1 + 0.055 (1 - a) \right]$
from the central values of this fit.
The uncertainties correspond only to the total uncertainties
of $\hat{d}_z(z, z_c)$.
To illustrate correlated uncertainties
we present results for
$(d_\textrm{BAO}, \Omega_\textrm{DE}) = (0.0339, 0.718 + 0.004)$ (squares),
$(0.0339, 0.718 - 0.004)$ (triangles),
$(0.0339 + 0.0003, 0.718)$ (inverted triangles), and
$(0.0339 - 0.0003, 0.718)$ (circles).
For clarity some offsets in $a$ have been applied.
}
\label{O_DE_z}
\end{center}
\end{figure}

\begin{figure}
\begin{center}
\scalebox{0.45}
{\includegraphics{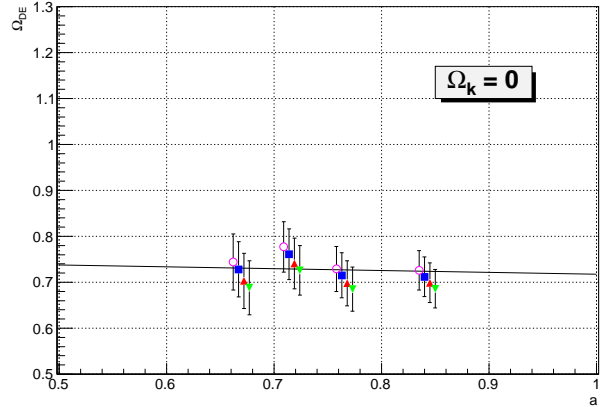}}
\caption{
Same as Fig. \ref{O_DE_z} for offset bins of $z$
with least $|Q - 1|$ in Tables \ref{N} or \ref{S}.
These measurements are partially correlated
with those of Fig. \ref{O_DE_z}.
}
\label{O_DE_z_offset}
\end{center}
\end{figure}

We obtain $\Omega_\textrm{DE}(a)$ from the 6 independent measurements of
$\hat{d}_z(z, z_c)$ in Table \ref{N_S},
and Eqs. (\ref{dz_rs}) and (\ref{E})
for the case $\Omega_k = 0$.
The values of $d_\textrm{BAO}$ and 
$\Omega_\textrm{m} = 1 - \Omega_\textrm{DE}(1) - \Omega_k$
are obtained from the fit for Scenario 4 in Table \ref{BAO_thetaMC_fit_18}.
The results are presented in Fig. \ref{O_DE_z}. To guide the eye, we
also show the straight line corresponding to 
the central values of $\Omega_\textrm{DE}$ and $w_1$ of 
the fit for Scenario 4. In Fig. \ref{O_DE_z_offset} we present
the results for offset bins of $z$ (which are partially correlated with
the entries in Fig. \ref{O_DE_z}).

\section{Measurements of $\Omega_k$, $\Omega_\textrm{DE}(a)$
and $N_\textrm{eq}$ from calibrated BAO}

Up to this point we have used the BAO distance
$d_\textrm{BAO} f = r_S$ as an uncalibrated standard ruler.
The cosmological parameters $h$ and $\Omega_\textrm{b} h^2$
drop out of such an analysis, and the dependences of the results
on $N_\textrm{eq}$ are not significant. 
$\Omega_\textrm{b} \equiv \rho_{b0}/\rho_\textrm{crit}$ 
is the present density of baryons relative to the critical density.
In this section we
consider the BAO distance as a calibrated standard ruler 
to constrain the cosmological parameters
$\Omega_k$, $\Omega_\textrm{DE}(a)$, $N_\textrm{eq}$,
$h$ and $\Omega_\textrm{b} h^2$.

The sound horizon is calculated from first principles \cite{Eisenstein}
as follows:
\begin{equation}
r'_S = \int_0^{t_\textrm{\tiny{dec}}} {\frac{c_s dt}{a}} = 
\int_0^{a_\textrm{\tiny{dec}}} \frac{c_s da}{H_0 a^2 E(a)},
\end{equation}
where the speed of sound is
\begin{equation}
c_s = \frac{c}{\sqrt{3 (1 + 3 \rho_{b0} a / (4 \rho_{\gamma 0}))}}.
\end{equation}
We can write the result for our purposes as 
\begin{equation}
r_S = 0.03389 \times A \times \left( \frac{0.30}{O_\textrm{m}} \right)^{0.255}
\end{equation} 
where
\begin{equation}
A = \left( \frac{h}{0.72} \right)^{0.489} 
  \left( \frac{0.023}{\Omega_\textrm{b} h^2} \right)^{0.098} 
  \left( \frac{3.36}{N_\textrm{eq}} \right)^{0.245}
\label{A}
\end{equation}
(we have neglected the dependence of $z_\textrm{dec} = 1089.9 \pm 0.4$ \cite{PDG, Planck} on 
the cosmological parameters). 

In this paragraph we take
$N_\textrm{eq} = 3.38$ corresponding
to 3 flavors of neutrinos \cite{PDG}.
From Big-Bang nucleosynthesis,
$\Omega_\textrm{b} h^2 = 0.0225 \pm 0.0008$ 
(at 68\% confidence) \cite{PDG}. 
With the latest direct measurement $h = 0.720 \pm 0.030$
by the Hubble Space Telescope Key Project \cite{hubble} we obtain $A = 1.000 \pm 0.021$.
An alternative choice is the Planck ``TT+lowP+lensing" analysis
\cite{PDG}, that assumes $\Omega_k = 0$ and a $\Lambda$CDM cosmology, that obtains
$\Omega_\textrm{b} h^2 = 0.02226 \pm 0.00023$, $h = 0.678 \pm 0.009$
and $A = 0.973 \pm 0.007$. 
The cosmological parameters that minimize the $\chi^2$ with
22 terms (18 BAO measurements from Table \ref{N_S} plus $\theta_\textrm{MC}$ from
Eq. (\ref{tMC}) plus 2 Ly$\alpha$ measurements from Eq. (\ref{Lyalpha})
plus $A$) are presented in Table \ref{calibrated_BAO}.
Note that the addition of the external constraint from $A$ slightly reduces the
uncertainties of $\Omega_k$ and $w_1$ if $N_\textrm{eq} = 3.38$ is fixed.
Note in Table \ref{calibrated_BAO} that the data is consistent
with the constraints $\Omega_k = 0$ and constant $\Omega_\textrm{DE}(a)$
for both values of $A$.

In this paragraph we let $N_\textrm{eq}$ be free.
We turn the problem around: from 
18 BAO measurements from Table \ref{N_S} plus $\theta_\textrm{MC}$ from
Eq. (\ref{tMC}) plus 2 Ly$\alpha$ measurements from Eq. (\ref{Lyalpha})
we constrain $A$. 
The results are $A = 0.965 \pm 0.014$ for free $\Omega_k$ and
$\Omega_\textrm{DE}(a)$ allowed to vary as in Scenario 4,
$A = 0.983 \pm 0.005$ for $\Omega_k = 0$ fixed and
$\Omega_\textrm{DE}(a)$	allowed	to vary	as in Scenario 4,
and $A = 0.9855 \pm 0.0012$ for $\Omega_k = 0$ fixed and constant $\Omega_\textrm{DE}(a)$.
For free $\Omega_k$,
$\Omega_\textrm{DE}(a)$ allowed to vary as in Scenario 4,
$\Omega_\textrm{b} h^2 = 0.0225 \pm 0.0008$, and
$h = 0.720 \pm 0.030$ we obtain $N_\textrm{eq} = 3.92 \pm 0.40$
corresponding to $N_\textrm{eff} = 4.2 \pm 0.9$ neutrino flavors.
For $\Omega_k = 0$ fixed, constant $\Omega_\textrm{DE}(a)$,
$\Omega_\textrm{b} h^2 = 0.02226 \pm 0.00023$, and $h = 0.678 \pm 0.009$
we obtain $N_\textrm{eq} = 3.20 \pm 0.09$
corresponding to $N_\textrm{eff} = 2.64 \pm 0.20$ neutrino flavors.

\begin{table*}
\caption{\label{calibrated_BAO}
Cosmological parameters obtained from the 18 BAO measurements in Table \ref{N_S}
plus $\theta_\textrm{MC}$ from Eq. (\ref{tMC}) plus 2 Ly$\alpha$ measurements in Eq. (\ref{Lyalpha}) 
plus $A$ in several scenarios.
Corrections for peculiar motions are given by
Eq. (\ref{correction}). $N_\textrm{eq} = 3.38$.
Scenario 1 has $\Omega_\textrm{DE}(a)$ constant.
Scenario 4 has $\Omega_\textrm{DE}(a) = \Omega_\textrm{DE} \left[1 + w_1 (1 - a)\right]$.
}
\begin{ruledtabular}
\begin{tabular}{c|cccccc}
   & Scenario 1 & Scenario 1 & Scenario 4 & Scenario 4 & Scenario 4 & Scenario 4 \\
$A$ & $1.000 \pm 0.021$ & $0.972 \pm 0.007$ & $1.000 \pm 0.021$ & $1.000 \pm 0.021$ 
  & $0.972 \pm 0.007$ & $0.972 \pm 0.007$ \\
\hline
$\Omega_k$ & $0$ fixed  & $0$ fixed & $0$ fixed & $-0.006 \pm 0.008$ & $0$ fixed 
  & $-0.009 \pm 0.005$ \\
$\Omega_\textrm{DE} + 2.2 \Omega_k$ & $0.719 \pm 0.003$ & $0.719 \pm 0.003$ & $0.718 \pm 0.004$
  & $0.718 \pm 0.004$ & $0.715 \pm 0.004$ & $0.718 \pm 0.004$ \\
$w_1$ & n.a. & n.a. & $0.05 \pm 0.14$ & $0.16 \pm 0.20$ & $0.19 \pm 0.13$ & $0.24 \pm 0.12$ \\
$100 d_\textrm{BAO}$ & $3.40 \pm 0.02$ & $3.39 \pm 0.02$ & $3.39 \pm 0.03$
  & $3.38 \pm 0.03$ & $3.36 \pm 0.03$ & $3.38 \pm 0.03$ \\
$\chi^2/$d.f. & $20.3/20$ & $23.8/20$ & $20.1/19$ & $19.6/18$ & $21.1/19$ & $17.8/18$ \\
\end{tabular}
\end{ruledtabular}
\end{table*}

\section{Comparison with previous measurements}

Let us compare the results obtained with
SDSS DR13 data with DR12 data.
The $\chi^2$ between Table \ref{N_S} and
Table III of Ref. \cite{bh1} is 44.8 for 18 degrees
of freedom.
The $\chi^2$ between Table \ref{N_S} and
Table III of Ref. \cite{bh2} is 25.9 for 17 degrees
of freedom.
The disagreement in both cases is due to the same two
entries in Tables III of Ref. \cite{bh1} or III of Ref. \cite{bh2} with
miss-fits converging on background fluctuations
instead of the BAO signal: $\hat{d}_\alpha(0.46, 3.79)$ and
$\hat{d}_/(0.55, 3.79)$. The fluctuation of
$\hat{d}_\alpha(0.46, 3.79)$ can be seen in
Table I for the northern galactic cap, but not
in Table II for the southern galactic cap.
Removing the two miss-fits from	the comparisons
obtains	$\chi^2/\textrm{d.f.} = 21.2/16$ and
$11.1/15$ respectively.

We compare Eq. (\ref{19_terms}) for DR13 data, with the
corresponding fits for DR12 data. From Table VIII of Ref. \cite{bh1}:
\begin{eqnarray}
\Omega_k & = & 0.043 \pm 0.041, \nonumber \\
\Omega_\textrm{DE} + 2.0 \Omega_\textrm{k} & = & 0.716 \pm 0.006, \nonumber \\
w_1 & = & -0.16 \pm 0.94.
\end{eqnarray}
From Table VII of Ref. \cite{bh2}:
\begin{eqnarray}
\Omega_k & = & 0.060 \pm 0.052, \nonumber \\
\Omega_\textrm{DE} + 2.0 \Omega_\textrm{k} & = & 0.717 \pm 0.007, \nonumber \\
w_1 & = & -0.86 \pm 1.26.
\end{eqnarray}
Note in Eq. (\ref{19_terms}) how the DR13 data has lowered the uncertainties.

The final consensus measurements of
SDSS DR12 data \cite{consensus}
are presented in Table \ref{dr12} (reproduced from Ref. \cite{bh2}
for completness). There is agreement with the
measurements of DR13 data in Table \ref{N_S}.
The notation of Ref. \cite{consensus}
is related to the notation of the present article
as follows:
\begin{eqnarray}
D_M \frac{r_\textrm{d,fid}}{r_\textrm{d}} & = &
 \frac{c}{H_0} \chi(z) [1 + \frac{1}{6} \Omega_k \chi^2(z)]
\frac{r_\textrm{d,fid}}{d'_\textrm{BAO}} \nonumber \\
 & = & r_\textrm{d,fid} \frac{z \exp{(-z/z_c)}}{\hat{d}_\alpha(z, z_c)}, \\
H \frac{r_\textrm{d}}{r_\textrm{d,fid}} & = &
 H_0 E(z) \frac{d'_\textrm{BAO}}{r_\textrm{d,fid}} \nonumber \\
 & = & \frac{c}{r_\textrm{d,fid}} \frac{\hat{d}_z(z, z_c)}{(1 - z/z_c) \exp{(-z/z_c)}},
\end{eqnarray}
where $r_\textrm{d,fid}	= 147.78$ Mpc and $H_0 = 67.6$ km s$^{-1}$ Mpc$^{-1}$.

\begin{table*}
\caption{\label{dr12}
Final consensus ``BAO$+$FS" measurements of the	SDSS DR12 data set \cite{consensus}
(uncertainties are statistical and systematic), 
and the	corresponding BAO parameters $\hat{d}_\alpha(z, z_c)$ and
$\hat{d}_z(z, z_c)$ with $z_c =	3.79$. These measurements include the
peculiar motion	corrections. 
}
\begin{ruledtabular}
\begin{tabular}{c|cc|cc}
$z$ & $D_M r_\textrm{d,fid} / r_\textrm{d}$ [Mpc] & $100 \hat{d}_\alpha(z, z_c)$ &
 $H r_\textrm{d} / r_\textrm{d,fid} [\textrm{km s}^{-1} \textrm{Mpc}^{-1}]$ &
 $100 \hat{d}_z(z, z_c)$ \\
\hline
0.38 & $1518 \pm 20 \pm 11$ & $3.346 \pm 0.050$ 
 & $81.5 \pm 1.7 \pm 0.9$ & $3.270 \pm 0.077$ \\
0.51 & $1977 \pm 23 \pm 14$ & $3.332 \pm 0.045$ 
 & $90.5 \pm 1.7 \pm 1.0$ & $3.375 \pm 0.074$ \\
0.61 & $2283 \pm 28 \pm 16$ & $3.362 \pm 0.047$ 
 & $97.3 \pm 1.8 \pm 1.1$ & $3.426 \pm 0.074$
\end{tabular}
\end{ruledtabular}
\end{table*}

\section{Conclusions}

(i) The main results of these studies are
the independent measured BAO observables
$\hat{d}_\alpha(z, z_c)$, $\hat{d}_z(z, z_c)$, and $\hat{d}_/(z, z_c)$
presented in Table \ref{N_S}.
It is difficult to distinguish the BAO signal from fluctuations
of the background.
To gain confidence in the results we have repeated the measurements
many times with different galaxy selections to obtain
different background fluctuations. 
Requiring successful fits for each of the three
independent observables, for each bin of $z$, allows 
the use of the consistency relation (\ref{Q}) to
discriminate against miss-fits on background
fluctuations instead of the BAO signal. The consistency parameter $Q$ also
allows quality control of the measurements, see
Figure \ref{fig_Q}.
Table \ref{N_S} for DR13 data supersedes the
corresponding tables in Refs. \cite{bh1} and \cite{bh2}
for DR12 data.

(ii) From the 18 BAO measurements in Table \ref{N_S}, and no other input, we obtain
\begin{eqnarray}
\Omega_k & = & 0.151 \pm 0.185, \nonumber \\
\Omega_\textrm{DE} + 0.5 \Omega_\textrm{k} & = & 0.732 \pm 0.052, \nonumber \\
w_1 & = & 0.33 \pm 0.70,
\end{eqnarray}
for $\Omega_\textrm{DE}(a)$ allowed to vary as in Scenario 4.
For $\Omega_\textrm{k} = 0$ and constant $\Omega_\textrm{DE}(a)$
we obtain $\Omega_\textrm{DE} = 0.716 \pm 0.014$,
which may
be compared to the independent Planck ``TT$+$lowP$+$lensing" result
(which assumes a $\Lambda$CDM cosmology
with $\Omega_k = 0$):
$\Omega_\textrm{DE} = 0.692 \pm 0.012$ \cite{PDG}.
Note that these two results are based on independent
cosmological measurements.
See Table \ref{BAO_fit} for fits
in several scenarios.

(iii) From 18 BAO measurements plus $\theta_{MC}$ from the CMB
we obtain 
\begin{eqnarray}
\Omega_k & = & -0.015 \pm 0.030, \nonumber \\
\Omega_\textrm{DE} + 2.2 \Omega_\textrm{k} & = & 0.717 \pm 0.004, \nonumber \\
w_1 & = & 0.37 \pm 0.61,
\end{eqnarray}
for $\Omega_\textrm{DE}(a)$ allowed to vary as in Scenario 4.
See Tables \ref{BAO_thetaMC_fit_18}
for fits in several scenarios. The cosmological parameters
$h$, $\Omega_\textrm{b} h^2$ and $N_\textrm{eq}$ drop out
of this analysis.
Imposing the constraints $\Omega_k = 0$ and constant $\Omega_\textrm{DE}(a)$
obtains $\Omega_\textrm{DE} = 0.719 \pm 0.003$.

(iv) Detailed measurements of $\Omega_\textrm{DE}(a)$
are presented in Figs. \ref{O_DE_z} and \ref{O_DE_z_offset}.

(v) From 18 BAO plus $\theta_\textrm{MC}$ plus 2 Ly$\alpha$ measurements
we obtain 
\begin{eqnarray}
\Omega_k & = & -0.013 \pm 0.009, \nonumber \\
\Omega_\textrm{DE} + 2.2 \Omega_\textrm{k} & = & 0.717 \pm 0.004, \nonumber \\
w_1 & = & 0.34 \pm 0.24, \nonumber \\
A & = & 0.965 \pm 0.014,
\end{eqnarray}
when $\Omega_\textrm{DE}(a)$ is allowed to vary as in Scenario 4.
Note the constraint on $A$ defined in Eq. (\ref{A}).
The corresponding constraint on $N_\textrm{eq}$ for
$\Omega_\textrm{b} h^2 = 0.0225 \pm 0.0008$, and
$h = 0.720 \pm 0.030$ is $N_\textrm{eq} = 3.92 \pm 0.40$
corresponding to $N_\textrm{eff} = 4.2 \pm 0.9$ neutrino flavors.

For $\Omega_k = 0$ and constant $\Omega_\textrm{DE}(a)$
we obtain $A = 0.9855 \pm 0.0012$.
The corresponding constraint on	$N_\textrm{eq}$ for
$\Omega_\textrm{b} h^2 = 0.02226 \pm 0.00023$, and $h = 0.678 \pm 0.009$
is $N_\textrm{eq} = 3.20 \pm 0.09$
corresponding to $N_\textrm{eff} = 2.64 \pm 0.20$ neutrino flavors.

(vi) From 18 BAO plus $\theta_\textrm{MC}$ plus 2 Ly$\alpha$ 
plus $A$ measurements with $N_\textrm{eq} = 3.38$ fixed
we obtain the results shown in Table \ref{calibrated_BAO}.
For $\Omega_\textrm{DE}(a)$ allowed to vary as in Scenario 4
and $A = 1.000 \pm 0.021$ we obtain
\begin{eqnarray}
\Omega_k & = & -0.006 \pm 0.008, \nonumber \\
\Omega_\textrm{DE} + 2.2 \Omega_\textrm{k} & = & 0.718 \pm 0.004, \nonumber \\
w_1 & = & 0.16 \pm 0.20.
\end{eqnarray}

(vii) For all data sets we obtain no tension 
with the constraints $\Omega_k = 0$ and constant $\Omega_\textrm{DE}(a)$.

The SDSS has brought the measurements of $\Omega_\textrm{DE}(a)$ with
free $\Omega_k$ to a new level of precision.

\section{Acknowledgment}
Funding for the Sloan Digital Sky Survey IV has been provided by
the Alfred P. Sloan Foundation, the U.S. Department of Energy Office of
Science, and the Participating Institutions. SDSS-IV acknowledges
support and resources from the Center for High-Performance Computing at
the University of Utah. The SDSS web site is www.sdss.org.

SDSS-IV is managed by the Astrophysical Research Consortium for the 
Participating Institutions of the SDSS Collaboration including the 
Brazilian Participation Group, the Carnegie Institution for Science, 
Carnegie Mellon University, the Chilean Participation Group, 
the French Participation Group, Harvard-Smithsonian Center for Astrophysics, 
Instituto de Astrof\'isica de Canarias, The Johns Hopkins University, 
Kavli Institute for the Physics and Mathematics of the Universe (IPMU) / 
University of Tokyo, Lawrence Berkeley National Laboratory, 
Leibniz Institut f\"ur Astrophysik Potsdam (AIP),  
Max-Planck-Institut f\"ur Astronomie (MPIA Heidelberg), 
Max-Planck-Institut f\"ur Astrophysik (MPA Garching), 
Max-Planck-Institut f\"ur Extraterrestrische Physik (MPE), 
National Astronomical Observatories of China, New Mexico State University, 
New York University, University of Notre Dame, 
Observat\'ario Nacional / MCTI, The Ohio State University, 
Pennsylvania State University, Shanghai Astronomical Observatory, 
United Kingdom Participation Group,
Universidad Nacional Aut\'onoma de M\'exico, University of Arizona, 
University of Colorado Boulder, University of Oxford, University of Portsmouth, 
University of Utah, University of Virginia, University of Washington, University of Wisconsin, 
Vanderbilt University, and Yale University.


\begin{thebibliography}{7}

\bibitem{Eisenstein} 
D. J. Eisenstein, H.-J. Seo, and M. White, ApJ, 664: 660-674 (2007).

\bibitem{PDG}
C. Patrignani et al. (Particle Data Group), Chin. Phys. C, 40, 100001 (2016). 

\bibitem{BAO1} 
Bruce A. Bassett and Ren\'{e}e Hlozek, arXiv:0910.5224 (2009).

\bibitem{BAO2} 
David H. Weinberg et.al., arXiv:1201.2434 (2013).

\bibitem{BH} 
B. Hoeneisen, arXiv:astro-ph/0009071 (2000).

\bibitem{Seo}
Hee-Jong Seo, et al., ApJ, 720, 1650 (2010).

\bibitem{DR13}
Franco D. Albareti, et al., SDSS Collaboration, arXiv:1608.02013 (2016).

\bibitem{bh1}
B. Hoeneisen, arXiv:1607.02424 (2016).

\bibitem{Planck}
Planck Collab. 2015 Results XIII, Astron. \& Astrophys.
submitted, arXiv:1502.01589v2.

\bibitem{lyman}
Andreu Font-Ribera et. al., J. Cosmology Astropart. Phys. 05, 027 (2014), arXiv:1311.1767.

\bibitem{lyman2}
Timoth\'{e}e Delubac, et al., arXiv:1404.1801v2 (2014).

\bibitem{bh2}
B. Hoeneisen, arXiv:1608.08486 (2016).

\bibitem{DR12}
Shadab Alam, et al. (SDSS-III), arXiv:1501.00963 (2015).

\bibitem{Peacock}
John A. Peacock, ``Cosmological Physics", 
Cambridge University Press (1999).



\bibitem{Chevallier}
M. Chevallier, D. Polarski
Int. J. Mod. Phys. D10, 213 (2001)

\bibitem{Linder}
Eric V. Linder, Phys.Rev.Lett. 90:091301 (2003)


\bibitem{hubble}
E.M.L. Humphreys et al., Astrophys. J., 775, 13 (2013).

\bibitem{consensus} 
Shadab Alam et al., arXiv:1607.03155 (2016).

\end{thebibliography}
\end{document}